\documentclass[aps,pra,reprint,superscriptaddress,twocolumn,longbibliography]{revtex4-1}
\usepackage{amsfonts,amssymb,amscd,amsthm}
\usepackage{graphicx}
\usepackage{mathrsfs}
\usepackage{soul,color,xcolor}
\usepackage[intlimits]{amsmath}
\usepackage[colorlinks, citecolor=red]{hyperref}
\usepackage{upgreek}
\usepackage{diagbox}

\begin{document}
	\title{Hardware-Economic Manipulation of Dual-Type ${}^{171}$Yb$^+$ Qubits}
	
	\author{Y.-J. Yi}
	\thanks{These authors contribute equally to this work}%
	\affiliation{Center for Quantum Information, Institute for Interdisciplinary Information Sciences, Tsinghua University, Beijing 100084, PR China}
	
	\author{Y.-Y. Chen}
	\thanks{These authors contribute equally to this work}%
	\affiliation{Center for Quantum Information, Institute for Interdisciplinary Information Sciences, Tsinghua University, Beijing 100084, PR China}
	
	\author{Y.-H. Hou}
	\thanks{These authors contribute equally to this work}%
	\affiliation{HYQ Co., Ltd., Beijing 100176, P. R. China}
	
	\author{Y.-K. Wu}
	
	\affiliation{Center for Quantum Information, Institute for Interdisciplinary Information Sciences, Tsinghua University, Beijing 100084, PR China}
	\affiliation{Hefei National Laboratory, Hefei 230088, PR China}
	
	\author{L. Zhang}
	\affiliation{Center for Quantum Information, Institute for Interdisciplinary Information Sciences, Tsinghua University, Beijing 100084, PR China}
	
	\author{C. Zhang}
	\affiliation{HYQ Co., Ltd., Beijing 100176, P. R. China}
	
	\author{Y.-L. Xu}
	\affiliation{Center for Quantum Information, Institute for Interdisciplinary Information Sciences, Tsinghua University, Beijing 100084, PR China}
	
	\author{J. Ye}
	\affiliation{Center for Quantum Information, Institute for Interdisciplinary Information Sciences, Tsinghua University, Beijing 100084, PR China}
	
	\author{W.-X. Guo}
	\affiliation{HYQ Co., Ltd., Beijing 100176, P. R. China}
	
	\author{B.-X. Qi}
	\affiliation{Center for Quantum Information, Institute for Interdisciplinary Information Sciences, Tsinghua University, Beijing 100084, PR China}
	%\affiliation{Hefei National Laboratory, Hefei 230088, PR China}
	
	\author{Z.-C. Zhou}
	\affiliation{Center for Quantum Information, Institute for Interdisciplinary Information Sciences, Tsinghua University, Beijing 100084, PR China}
	\affiliation{Hefei National Laboratory, Hefei 230088, PR China}
	
	\author{P.-Y. Hou}
	\affiliation{Center for Quantum Information, Institute for Interdisciplinary Information Sciences, Tsinghua University, Beijing 100084, PR China}
	\affiliation{Hefei National Laboratory, Hefei 230088, PR China}
	
	\author{L.-M. Duan}
	\email{lmduan@tsinghua.edu.cn}
	\affiliation{Center for Quantum Information, Institute for Interdisciplinary Information Sciences, Tsinghua University, Beijing 100084, PR China}
	\affiliation{Hefei National Laboratory, Hefei 230088, PR China}
	
	\begin{abstract}
		The dual-type qubit scheme is an emerging method to suppress crosstalk errors in scalable trapped-ion quantum computation and quantum network.
		Here we report a hardware-economic way to control dual-type $^{171}\mathrm{Yb}^+$ qubits using a single $355\,$nm mode-locked pulsed laser. Utilizing its broad frequency comb structure, we drive the Raman transitions of both qubit types encoded in the $S_{1/2}$ and the $F_{7/2}$ hyperfine levels, and probe their carrier transitions and the motional sidebands. We further demonstrate a direct entangling gate between the two qubit types. Our work can simplify the manipulation of the $^{171}\mathrm{Yb}^+$ qubits both at the hardware and the software level.
	\end{abstract}
	
	\maketitle
	
	The dual-type qubit scheme has become a promising approach toward large-scale trapped-ion quantum computation and quantum network \cite{yang2022realizing,10.1063/5.0069544}. By encoding different qubit types into the ground-state and the metastable-state manifolds of the same ion species with distinct transition frequencies, their crosstalk error can be avoided which may arise from the random photon scattering during the dissipative operations. For example, it has been applied for sympathetic laser cooling, state initialization and readout of ancilla qubits \cite{yang2022realizing,PhysRevLett.132.263201} and for the repetitive ion-photon entanglement generation on the communication qubits \cite{Feng2024,PhysRevLett.134.070801} with negligible crosstalk error on the data qubits storing quantum information. Recently, a storage lifetime above two hours has been achieved for the memory ions under the continuous sympathetic cooling of the ancilla ions using the dual-type scheme \cite{mjqd-9mvf}. Compared with the dual-species approach where the two qubit types are encoded into different ion species or different isotopes \cite{PhysRevA.65.040304,PhysRevA.68.042302,PhysRevA.79.050305,guggemos2015sympathetic,PhysRevLett.118.250502,negnevitsky2018repeated,PhysRevLett.130.090803}, the dual-type scheme has the advantage that the two types can be coherently converted into each other on demand, allowing arbitrary assignment of their proportions and locations in a large ion crystal without the need of the relatively slow ion transport operations \cite{Kielpinski2002,Pino2021}. Besides, the two qubit types share the same mass, which is beneficial for the efficiency of sympathetic laser cooling \cite{PhysRevA.103.012610}.
	
	While the large frequency difference between the two qubit types is critical for the suppression of the crosstalk error, it also means that typically they need to be manipulated by different laser setups \cite{quinn2024high}. A workaround is to always bring the qubits into a certain type for manipulations, and then back to the original type \cite{yang2022realizing,10.1063/5.0069544}. Nevertheless, either method amounts to an additional cost for the dual-type qubit scheme, whether at the hardware level or at the software level. To solve this problem, recently a hardware-economic scheme has been demonstrated for the $^{137}\mathrm{Ba}^+$ ions where two qubit types encoded in both $S_{1/2}$ and $D_{5/2}$ manifolds can be manipulated by $532\,$nm Raman transitions, and a direct entangling gate between the two qubit types has been achieved \cite{PhysRevLett.134.010601}. However, the success of such a scheme largely depends on the existence of suitable intermediate energy levels, and whether it can be generalized to other commonly used ion species for quantum information processing like $^{171}\mathrm{Yb}^+$ is not clear.
	
	In this work, we report an integrated control scheme for dual-type $^{171}\mathrm{Yb}^+$ qubits encoded in the $S_{1/2}$ and $F_{7/2}$ manifolds. We utilize the broad frequency comb structure of a single mode-locked $355\,$nm pulsed laser \cite{Debnath2016,debnath2016programmable} to drive the Raman transition of both qubit types. We drive the carrier transitions and the motional sidebands for a pair of $S$-type and $F$-type qubits, and we demonstrate a direct entangling gate between them.

	\begin{figure*}[!tbp]
		\centering
		\includegraphics[width=\linewidth]{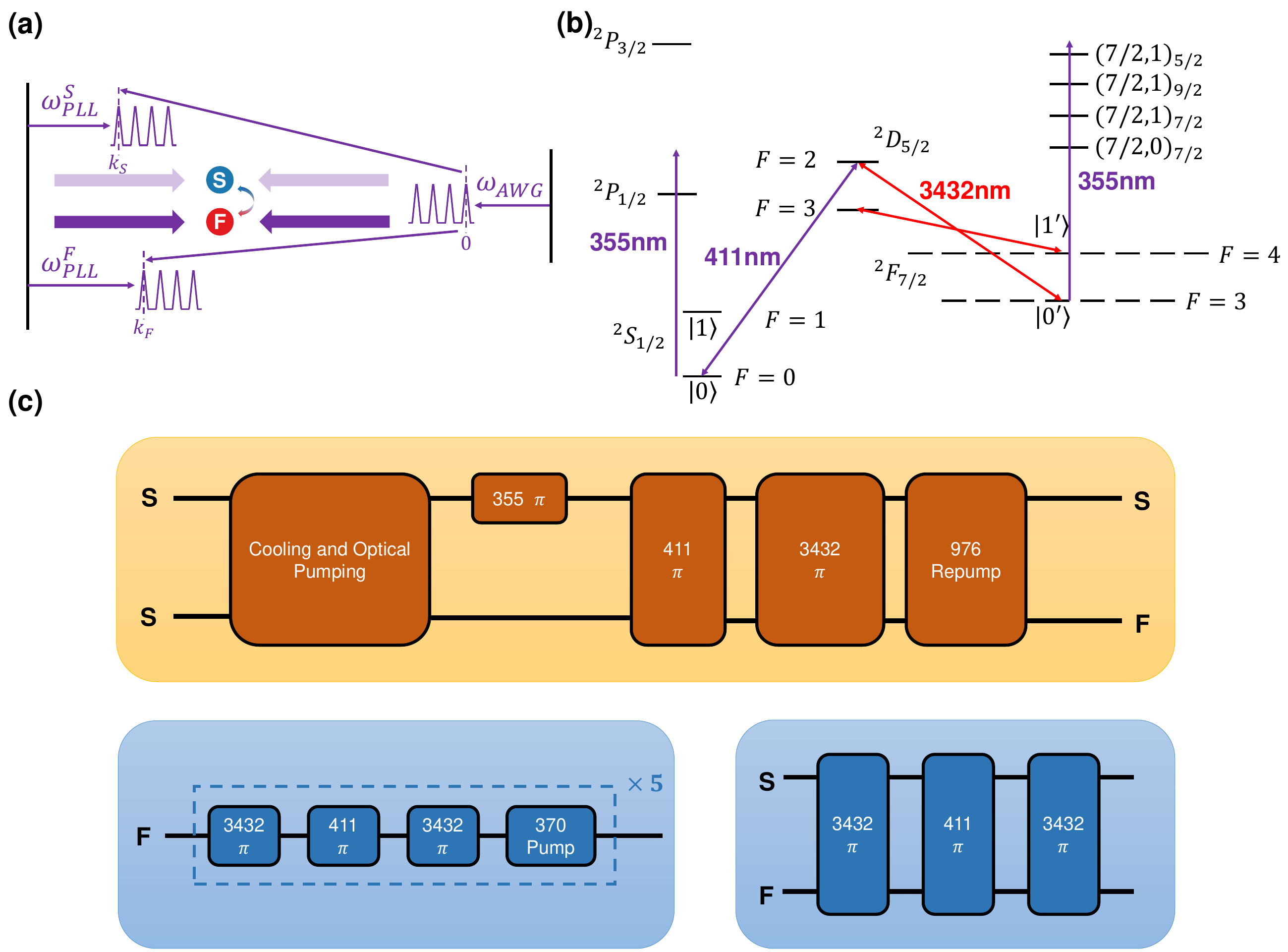}
		\caption {Experimental scheme. (a) Counter-propagating $355\,$nm pulsed laser beams with a repetition rate $\omega_r=2\pi\times 80\,$MHz control both the $S$-type and the $F$-type qubits, whose Raman transitions are bridged by the $k_S=158$ and the $k_F=45$ teeth of the frequency-comb structure, respectively. The laser beams from one direction are locked to the transition frequencies of the two qubit types by phase-locked loops (PLLs) to compensate the drift of $\omega_r$. The laser beams in the other direction are tuned by an arbitrary waveform generator (AWG) to select the carrier transitions or the blue and the red motional sidebands.
			(b) Relevant energy levels of $^{171}\mathrm{Yb}^+$. We use $411\,$nm and $3432\,$nm lasers to connect the $S$-type qubit spanned by $|0\rangle$ and $|1\rangle$, and the $F$-type qubit spanned by $|0^\prime\rangle$ and $|1^\prime\rangle$. We use $355\,$nm laser to drive Raman transitions of both qubit types, mediated by the upper energy levels.
			(c) Experimental sequence. The upper panel shows the initialization of an $S$-$F$ ion pair. The lower panel shows the detection of the $F$-type qubit (left) and the simultaneous detection of $S$-type and $F$-type qubits (right). The detection sequences are followed by fluorescence collection under global $370\,$nm laser beams which are not shown in the plot.} \label{fig1}
	\end{figure*}
	
	Our experimental scheme is shown in Fig.~\ref{fig1}. We encode dual-type qubits in the $S_{1/2}$ and $F_{7/2}$ hyperfine levels of $^{171}\mathrm{Yb}^+$ as $|0\rangle \equiv |S_{1/2},F=0,m_s=0\rangle$, $|1\rangle \equiv |S_{1/2},F=1,m_s=0\rangle$, $|0^\prime\rangle\equiv|F_{7/2},F=3,m_s=0\rangle$ and $|1^\prime\rangle\equiv|F_{7/2},F=4,m_s=0\rangle$.
	With the help of the upper energy levels as sketched in Fig.~\ref{fig1}(b) \cite{roman_expanding_2021}, we can use $355\,$nm laser to drive Raman transitions of both the $S$-type and the $F$-type qubits. Furthermore, because we use a pulsed laser with a pulse width of $15\,$ps and a repetition rate of $\omega_r=2\pi\times 80\,$MHz, it has a broad bandwidth of $67\,$GHz in its frequency comb structure \cite{PhysRevLett.104.140501,Debnath2016,debnath2016programmable} which covers the hyperfine splitting of $\omega_S=2\pi\times 12.642\,$GHz for the $S$-type qubit and $\omega_F=2\pi\times 3.620\,$GHz for the $F$-type qubit. Therefore, it is sufficient to use acousto-optic modulators (AOMs) to bridge their transitions together with $k_S=158$ and $k_F=45$ teeth of the frequency comb. As shown in Fig.~\ref{fig1}(a), for the $355\,$nm laser beams from one direction, we have an AOM controlled by phase-locked loops (PLLs) at the frequencies of $\omega_{\mathrm{PLL}}^S$ and $\omega_{\mathrm{PLL}}^F$ to stabilize the Raman transition frequencies to the two qubit types \cite{Islam:14} (see Appendix~\ref{app:setup} for more details), while for the laser beam from the other direction we use an additional AOM controlled by an arbitrary waveform generator (AWG) at the frequency of $\omega_{\mathrm{AWG}}$ to select the detuning $\delta$ of the Raman beams from the carrier transitions. Specifically we have
	\begin{equation}
		\omega_{\mathrm{AWG}} - \left[\omega_{\mathrm{PLL}}^{S(F)} - k_{S(F)} \omega_r\right] = \omega_{S(F)} + \delta. \label{eq1}
	\end{equation}
	In principle, we can set a shared $\omega_{\mathrm{AWG}}$ to control both qubit types. However, in practice due to the finite bandwidth of the AOMs, we choose to set $\omega_{\mathrm{PLL}}^{S}=2\pi\times 240\,$MHz and the corresponding $\omega_{\mathrm{AWG}}$ around $2\pi\times 242\,$MHz for the $S$-type qubit, and set $\omega_{\mathrm{PLL}}^{F}=2\pi\times 250\,$MHz and the corresponding $\omega_{\mathrm{AWG}}$ around $2\pi\times 230\,$MHz for the $F$-type qubit.
	
\begin{figure}[!tbp]
		\centering
		\includegraphics[width=\linewidth]{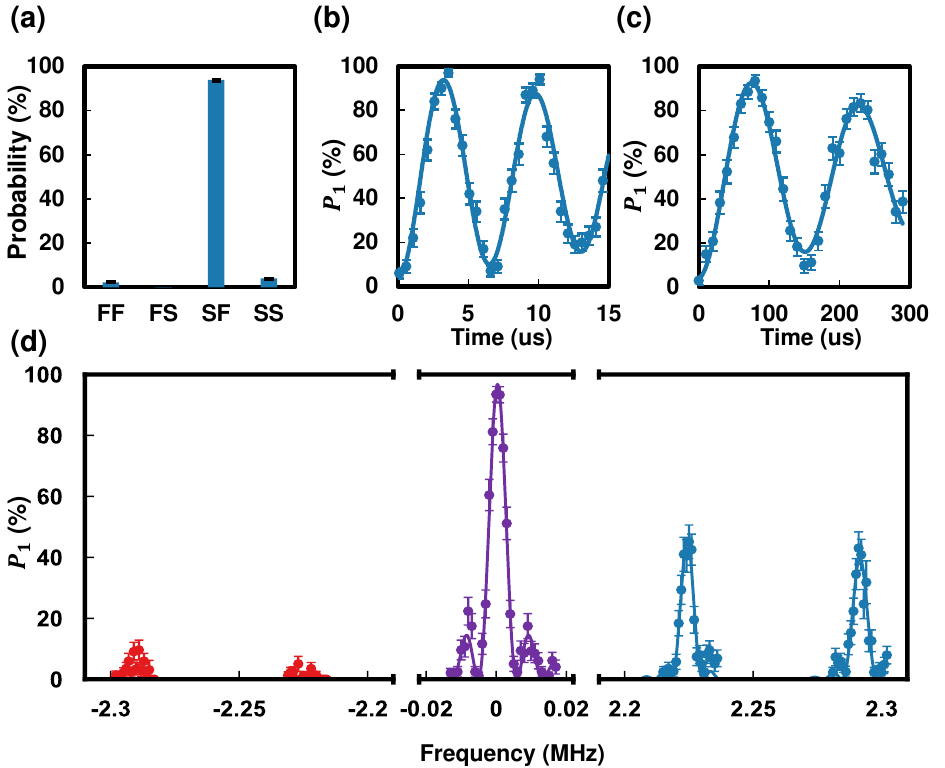}
		\caption {Individual manipulation of $S$-type and $F$-type qubits. (a) Success probability of one attempt to prepare the $S$-$F$ ion pair. (b) and (c) Carrier Rabi oscillations of the $S$-type and the $F$-type qubits under the same intensity of the $355\,$nm laser.
			(d) Carrier and motional sideband spectra of the $F$-type qubit after sympathetic sideband cooling on the $S$-type qubit.
		} \label{fig2}
	\end{figure}

	To demonstrate the manipulation of the two qubit types, first we prepare a pair of $S$-type and $F$-type qubits. The individual addressing is achieved by a pair of symmetrically placed acousto-optic deflectors (AODs) as in our previous work \cite{hou_individually_2024}, which guide the focused counter-propagating $355\,$nm laser beams with a radius of about $1.7 \,\upmu$m to the target ions.
	As shown in the upper panel of Fig.~\ref{fig1}(c), we start from two ions in $|0\rangle$ by $370\,$nm Doppler cooling and optical pumping, and perform an individual $355\,$nm $\pi$ pulse to flip the first qubit into $|1\rangle$ with a crosstalk error below $0.1\%$ on the second ion at a distance of about $4.5\,\upmu$m \cite{hou_individually_2024}. Then the second ion is mapped into $|0^\prime\rangle$ by global $411\,$nm and $3432\,$nm $\pi$ pulses following the transition paths in Fig.~\ref{fig1}(b). In this experiment, our global $411\,$nm laser selectively couples $|0\rangle$ to the $|D_{5/2},F=2,m_s=0\rangle$ level without affecting the first qubit in the $|1\rangle$ state. On the other hand, the global $3432\,$nm laser is bichromatic and couples both $|D_{5/2},F=2,m_s=0\rangle \leftrightarrow |0^\prime\rangle$ and $|D_{5/2},F=3,m_s=0\rangle \leftrightarrow |1^\prime\rangle$ transitions, because it is shared with other setups in our group for the coherent conversion of the dual-type qubits \cite{yang2022realizing,Feng2024}. Because of the imperfection in the laser $\pi$ pulses, we further add a verification step for the state preparation. We apply a global $976\,$nm laser to repump the population in $D_{5/2}$ back to $S_{1/2}$, followed by the fluorescence detection under global $370\,$nm laser. A successful initialization of the two qubits is thus indicated by the first ion being bright and the second ion being dark, with a success probability of about $94\%$ as illustrated in Fig.~\ref{fig2}(a). Otherwise we decide this attempt of state preparation to be failed and reset the $F$-type qubit into the $S$-type by $3432\,$nm and $976\,$nm laser beams for the next attempt.
	
	The above sequence of $411\,$nm and $3432\,$nm $\pi$ pulses can also be regarded as a detection sequence to distinguish $|0\rangle$ and $|1\rangle$ states for the $S$-type qubit. Similarly, to distinguish the $|0^\prime\rangle$ and $|1^\prime\rangle$ states for the $F$-type qubit, we use the experimental sequence in the lower left panel of Fig.~\ref{fig1}(c). Ideally, the first $3432\,$nm $\pi$ pulse and the $411\,$nm $\pi$ pulse act as a reverse process to convert $|0^\prime\rangle$ into $|0\rangle$, while the second $3432\,$nm $\pi$ pulse cancels the undesired effect of the bichromatic $3432\,$nm laser and brings the $|1^\prime\rangle$ state, which has been temporarily moved to $D_{5/2}$, back to the $F_{7/2}$ manifold. In practice, considering the imperfection of the pulses (in particular the $411\,$nm $\pi$ pulse), we further apply a $370\,$nm laser with the $935\,$nm repump laser turned off. This will pump all the population in $S_{1/2}$ to the $D_{3/2}$ levels, and will allow us to execute additional rounds of electron shelving from the $F_{7/2}$ levels. By repeating this sequence for five times, we then collect the fluorescence under a $370\,$nm detection laser and obtain a detection infidelity of $0.5\%$ for the $|0^\prime\rangle$ state and $0.9\%$ for the $|1^\prime\rangle$ state.
	
	To detect the states of the $S$-type and the $F$-type qubits simultaneously, we consider the experimental sequence in the lower right panel of Fig.~\ref{fig1}(c), which is identical to the initial part of the previous sequence for the $F$-type qubit alone. It leaves $|1\rangle$ and $|1^\prime\rangle$ untouched but exchanges the population of $|0\rangle$ and $|0^\prime\rangle$. Therefore, given the information of a qubit being in the $S$-type ($F$-type), we can decide its state as $|0\rangle$ or $|1\rangle$ ($|1^\prime\rangle$ or $|0^\prime\rangle$) depending on whether it is dark or bright under the $370\,$nm detection laser. Because of the pulse error, this time we have a detection infidelity of about $0.5\%$ for the $S$-type qubit and $2.8\%$ for the $F$-type qubit. In this experiment, when the simultaneous detection of both qubit types are necessary, such as calibrating the fidelity of a direct $S$-$F$ entangling gate, we further use the maximum likelihood method to compensate the detection error \cite{hou_individually_2024,Eliason_1993} (see Appendix~\ref{app:MLE} for more details).
	
	Next we demonstrate the manipulation of individual $S$-type and $F$-type qubits. In Fig.~\ref{fig2}(b) and (c) we drive the carrier Rabi oscillation of the $S$-type and the $F$-type qubits, respectively, using the same intensity of the $355\,$nm Raman laser beams. Typically, the Rabi rate for the $S$-type qubit is about 10 to 20 times of that for the $F$-type qubit, which fluctuates together with the central wavelength of the $355\,$nm laser on the timescale of a few days. On the other hand, the decay of the oscillation amplitude for the $F$-type qubit is much slower, as the decay per oscillation period is only about twice of that for the $S$-type qubit. This suggests that the dominant error source is the fluctuation of the $355\,$nm laser intensity which leads to a decay rate proportional to the Rabi rate. We can further drive the motional sidebands on both qubit types. As shown in Fig.~\ref{fig2}(d), for the two collective phonon modes at the frequencies of $2\pi\times 2.225\,$MHz and $2\pi\times 2.290\,$MHz, we perform the resolved Raman sideband cooling on the $S$-type qubit \cite{RevModPhys.75.281}, and probe the carrier and the sideband transitions on the $F$-type qubit.
	By comparing the strength of the red and the blue motional sidebands, we estimate an average phonon number of $\overline{n}_c=0.3$ for the center-of-mass mode, and $\overline{n}_r=0.1$ for the rocking mode.
	
	\begin{figure}[!tbp]
		\centering
		\includegraphics[width=\linewidth]{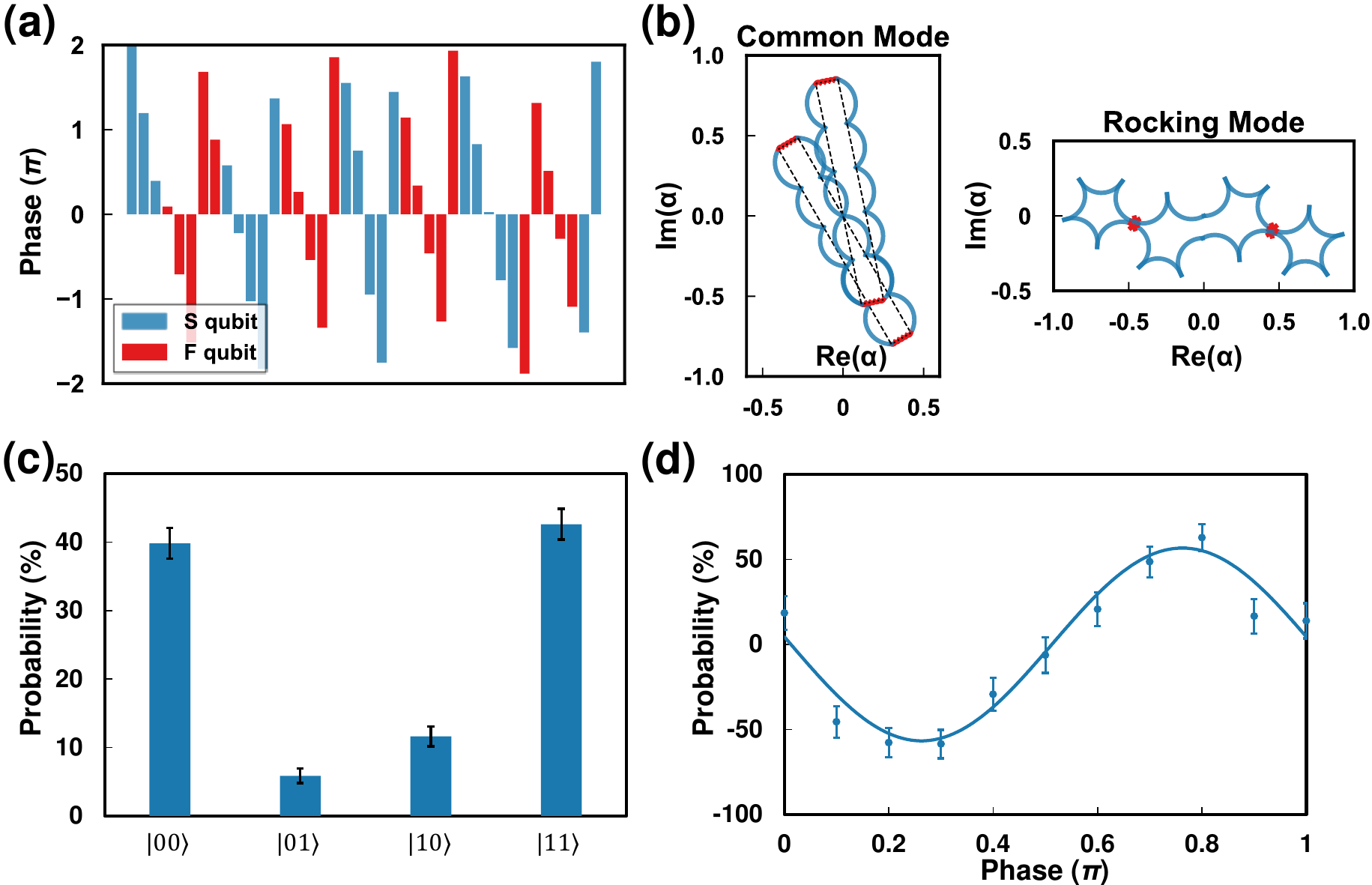}
		\caption {Direct entanglement of an $S$-$F$ qubit pair. (a) The blue (bright) and red (dark) bars indicate the phase modulation sequence alternatingly applied on the $S$-type and $F$-type qubits, respectively.
			(b) Theoretical phase space trajectories of the two collective phonon modes when the qubits are in $|+ +^\prime\rangle$.
			(c) Population and (d) parity oscillation of the prepared Bell state of the $S$-$F$ pair.
		} \label{fig3}
	\end{figure}
	
	Finally, we address the two qubit types simultaneously and perform a direct entangling gate between them.
	As described below Eq.~(\ref{eq1}), the two qubit types require different frequency settings on the AOMs controlling the counter-propagating laser beams. In principle we can apply those frequencies on the AOMs at the same time, and expect only small crosstalk errors due to the large off-resonance, but the cost will be a lowered laser intensity for each component because of the nonlinear response of the AOMs. Therefore, we utilize the alternating gate scheme \cite{hou_individually_2024} to quickly switch the addressing laser between the two target ions during the entangling gate, such that at any moment only a single frequency setup is needed for the AOMs. We then generate the spin-dependent force in a phase-sensitive geometry \cite{Lee_2005} with the AWG generating symmetric driving around the blue and the red motional sidebands simultaneously, and disentangle both collective phonon modes at the end of the gate by phase modulation \cite{PhysRevLett.114.120502}. We set the laser detuning in the middle of the two phonon modes, such that in this simple case the phase modulation sequence can be designed analytically. Specifically, we employ a heuristic sequence in Fig.~\ref{fig3}(a) with a ``rectangular'' phase space trajectory for the center-of-mass mode as shown in Fig.~\ref{fig3}(b). We divide the total gate time of $T=470.5\,\upmu$s into $n=40$ segments, each with a duration of $4\pi/[3(\omega_c-\omega_r)]\approx 9.81\,\upmu$s (here the mode frequencies are slightly shifted to $\omega_c=2\pi\times 2.271\,$MHz and $\omega_r=2\pi\times 2.203\,$MHz) and a separation of $2\,\upmu$s to switch the addressing signals on the AOMs and AODs.
	Each edge of the ``rectangle'' consists of five sequential segments acting on the same qubit type, after which we switch to the other qubit type for the next edge. The first edge starts from the $S$-type qubit, and we move two of its five segments to the end of the phase modulation sequence such that the overall shape of the phase space trajectory is maintained while its center is shifted closer to the origin, thus suppressing the error due to too large phonon excitations. In this way, we disentangle the center-of-mass mode using $20$ phase modulation segments, and we repeat this sequence again and adjust the relative phase between them to disentangle the rocking mode.
	
	To calibrate the fidelity of the entangling gate, we measure the population and the parity oscillation of the prepared Bell state \cite{debnath2016programmable} in Fig.~\ref{fig3}(c) and (d). As mentioned above, we use the maximum likelihood method to correct the detection infidelity (see Appendix~\ref{app:MLE}). From the measured population of $(82.5\pm 1.7)\%$ and the parity contrast of $(57\pm5)\%$, we estimate the Bell state fidelity of $F=(70\pm 3)\%$. This performance is mainly limited by the spin coherence time of $2.5\,$ms and the motional coherence time of $2\,$ms, as the weaker coupling to the $F$-type qubit leads to much slower $S$-$F$ entangling gate than our previous $S$-$S$ gate. Nevertheless, we note that these are not fundamental limitations. The spin dephasing mainly comes from the fluctuation in the optical path of the counter-propagating $355\,$nm laser beams, and can be improved by better stabilization of the optical system, while the motional dephasing can be suppressed by better locking of the trap frequency. Besides, it is also possible to increase the laser intensity when addressing the $F$-type qubits to shorten the gate duration. Note that in the experiment we occasionally observe the second ionization of the $F$-type ion under strong $355\,$nm laser driving (see Appendix~\ref{app:ionization} for more details). Further study will be needed to optimize the central wavelength and the spectrum of the $355\,$nm laser to speed up the Raman transition while maintaining low probability of the second ionization.
	
	To sum up, in this work we demonstrate the capability of controlling the $S$-type and $F$-type qubits using a single $355\,$nm pulsed laser. We drive the carrier and the motional sideband transitions of individual qubit types, and achieve a direct entangling gate between the two qubit types. Currently due to the lower coupling strength to the $F$-type qubit, the $S$-$F$ gate takes longer time and has larger gate infidelity than the $S$-$S$ gate \cite{hou_individually_2024}, so that performing this gate directly may not be advantageous compared with first converting the qubits back to the $S$-type. However, the gate performance may be improved in the future by extending the spin and the motional coherence time. Besides, currently for the coherent conversion between the $S$-type and the $F$-type qubits of ${}^{171}\mathrm{Yb}^+$ ions, the $3432\,$nm laser is used which is difficult to focus to individual ions. Although it is still possible to achieve individual control by focused $411\,$nm laser beams using a ``global $3432\,$nm $\pi$ pulse''-``individual $411\,$nm $\pi$ pulse''-``global $3432\,$nm $\pi$ pulse'' sequence similar to the lower right panel of Fig.~\ref{fig1}(c) \cite{Feng2024}, the error due to imperfect global $3432\,$nm pulses may accumulate with the ion number. In such cases, direct manipulation of the $F$-type qubits using $355\,$nm laser beams can still be preferable.
	
	\begin{acknowledgments}
		This work was supported by Quantum Science and Technology-National Science and Technology Major Project (Grant No. 2021ZD0301601), Beijing Science and Technology Planning Project (Grant No. Z25110100040000), the National Natural Science Foundation of China (Grant No. 12575021 and 12574541), Tsinghua University Initiative Scientific Research Program, and the Ministry of Education of China. L.M.D. acknowledges in addition support from the New Cornerstone Science Foundation through the New Cornerstone Investigator Program. Y.K.W. and P.Y.H. acknowledge in addition support from the Dushi program from Tsinghua University.
	\end{acknowledgments}
	
	\section*{Data Availability}
	The data that support the findings of this article are openly available\cite{Yi2026Data}.
	
	\appendix
	\section{Experimental Setup}
	\label{app:setup}
\begin{figure*}[!tbp]
		\centering
		\includegraphics[width=\linewidth]{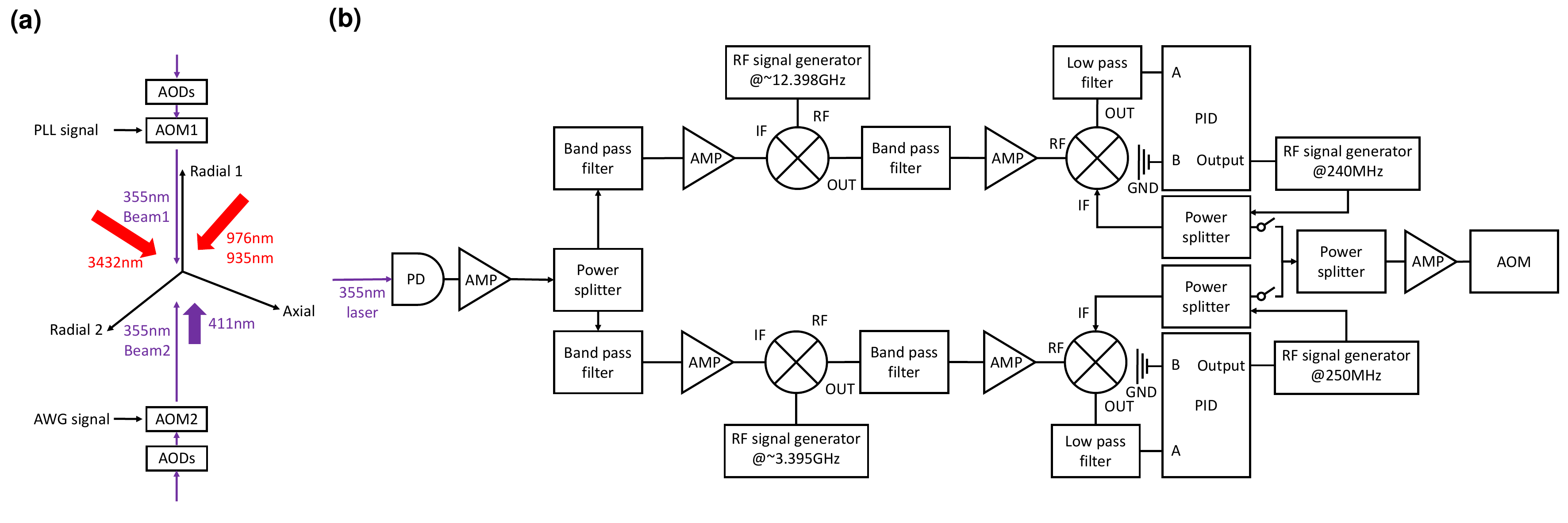}
		\caption{(a) Schematic of the laser setup.
			(b) Electronic system of the phase-locked loops for the $S$-type (upper) and the $F$-type (lower) qubits.
		} \label{supfig1}
	\end{figure*}

	The laser setup in our experiment is plotted in Fig.~\ref{supfig1}(a). Two counter-propagating $355\,$nm laser beams ($355\,$nm Beam $1$ and $355\,$nm Beam $2$) are directed to the ions along one of the two radial directions of the trap. The $411\,$nm laser is combined with $355\,$nm Beam~2 via a dichroic mirror. The $3432\,$nm laser enters the vacuum cavity through a side window. We also combine the repumping $935\,$nm and $976\,$nm laser into a co-propagating beam. In the plot, we omit the $370\,$nm laser beam for simplicity, which is at an angle to all the three principal axes of the trap for Doppler cooling, optical pumping and state detection.
	
	To lock the frequency of the $355\,$nm Raman transition to the qubit frequencies, we use the PLL electronic system in Fig.~\ref{supfig1}(b). A photodetector (PD) receives an incoming $355\,$nm laser beam and splits the amplified signal into two parts for the $S$-type and the $F$-type qubits. The signals go through band pass filters and keep only the frequency components near $k_S\omega_r=12.64\,$GHz and $k_F\omega_r=3.6\,$GHz, respectively. Then the signals are mixed with the outputs of two radiofrequency (RF) signal generators to obtain the difference frequency signals, and are further filtered by band pass filters to keep the difference frequency signals near $240\,$MHz and $250\,$MHz, respectively. These filtered signals are then mixed with $240\,$MHz and $250\,$MHz signals from two signal generators, respectively, to serve as the error signals for the PID servos. Then the outputs of the PID servos are used for the frequency modulation of the RF signal generators. Finally, the modulated signals from both paths are combined using a power splitter to drive the AOM. To implement the alternating gate scheme in the main text, we use electronic switches to ensure that only one PLL path is selected at any given time to drive either the $S$-type or the $F$-type qubit.
	
	\section{Correction of Detection Error}
	\label{app:MLE}
	To calibrate the fidelity of the direct $S$-$F$ entangling gate, we use the experimental sequence in the lower-right panel of Fig.~\ref{fig1}(c) to measure the two qubit types simultaneously, which suffers from the infidelity of the $411\,$nm and the $3432\,$nm $\pi$ pulses. To correct such detection errors and to focus on the performance of the entangling gate, we use the maximum likelihood method to recover the distribution of the measurement outcomes \cite{hou_individually_2024,Eliason_1993}.
	
	To calibrate the mapping between the input and the output distributions, we perform the preparation-detection sequence for about $10000$ times [about $6\%$ data are discarded due to the failure in the verification stage as described in Fig.~\ref{fig2}(a)] for each computational basis state $|00^\prime\rangle$, $|01^\prime\rangle$, $|10^\prime\rangle$, and $|11^\prime\rangle$. We use microwave $\pi$ pulses to prepare the $|1\rangle$ and $|1^\prime\rangle$ states, so that the state preparation error can be neglected compared with the detection error. In this way, we obtain the $4\times 4$ matrix $M$ as shown in Table~\ref{suptable1}. Then for any experimentally measured outcome distribution $\boldsymbol{f}=[f_0,f_1,f_2,f_3]^T$, we can numerically optimize the most likely probability distribution $\boldsymbol{p}=[p_0,p_1,p_2,p_3]^T$ such that the observed distribution $\boldsymbol{f}$ can be generated from the ideal distribution $M\boldsymbol{p}$ with the highest probability. With this recovered distribution $\boldsymbol{p}$, we can compute all the population and parity in Fig.~\ref{fig3} of the main text.
	
	\begin{table}[htbp]
		\centering
		\caption{Measurement outcomes for different input states. Probability values are presented in percentages.}
		\begin{tabular}{|c|c|c|c|c|}
			\hline
			\diagbox{\textbf{Measured}}{\textbf{Prepared}} & $|00'\rangle$ & $|01'\rangle$ & $|10'\rangle$ & $|11'\rangle$ \\
			\hline
			$|00'\rangle$ & 96.82 & 2.14 & 0.24 & 0.00    \\
			\hline
			$|01'\rangle$ & 2.27  & 97.16& 0.00 & 0.24 \\
			\hline
			$|10'\rangle$ & 0.91  & 0.23 & 97.43& 4.39 \\
			\hline
			$|11'\rangle$ & 0.00  & 0.47 & 2.33 & 95.38\\
			\hline
		\end{tabular}
		\label{suptable1}
	\end{table}
	
	\section{Second Ionization events}
	\label{app:ionization}
	\begin{figure}[!tbp]
		\centering
		\includegraphics[width=0.6\linewidth]{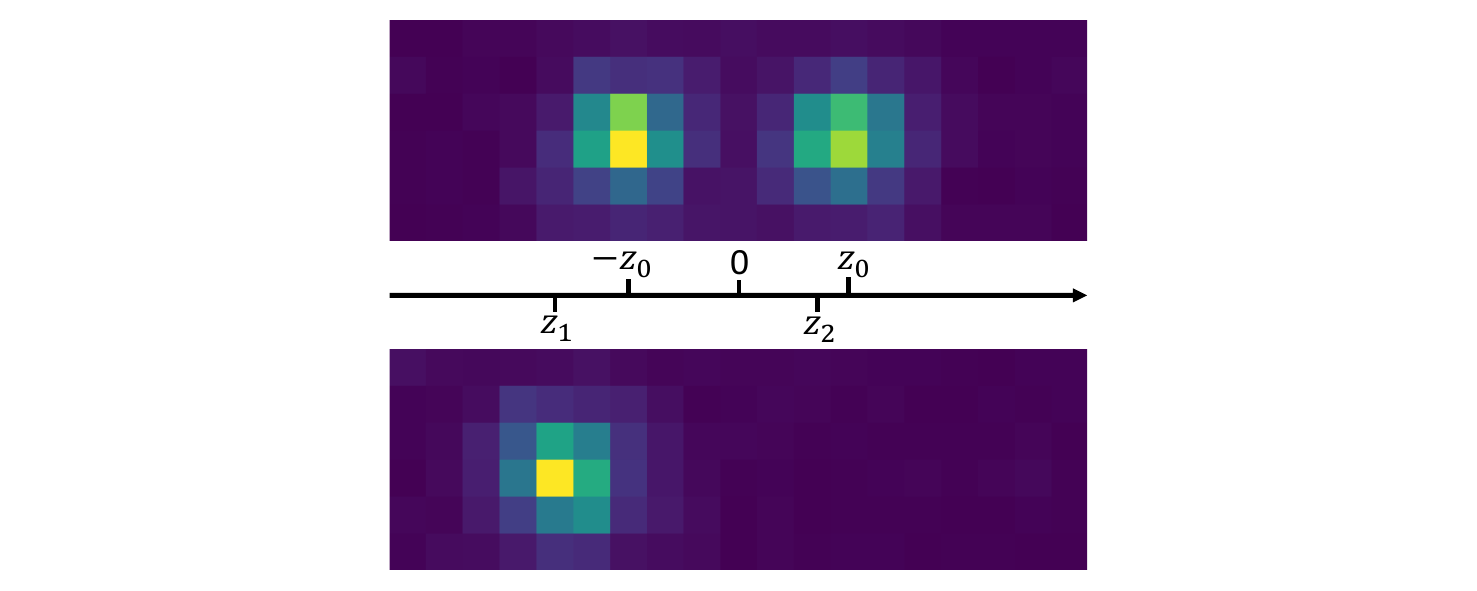}
		\caption{Positions of two ions before and after the second ionization.
		} \label{supfig2}
	\end{figure}
	In the experiment we occasionally observe second ionization of the ${}^{171}\mathrm{Yb}^+$ ion when applying strong $355\,$nm laser beams on the $F$-type qubit. This has also been observed previously in Ref.~\cite{johnson_experiments_2016}. Such an event is indicated in Fig.~\ref{supfig2}, where originally we have two monovalent $S$-type ions at the symmetric positions $\pm z_0$, while after converting the ion on the right to the $F$-type and applying the $355\,$nm laser beams for some time, they move to the new positions $z_1$ and $z_2$. The measured new location $z_1=-1.61 z_0$ is close to the theoretical result $z_1=-1.53 z_0$, distinguishing such events from the formation of an $\mathrm{YbH}^+$ ion.
	
	To speed up the manipulation of the $F$-type qubit without causing loss errors, it will be necessary to suppress the second ionization rate. The detailed transition path for the second ionization is not clear to us. However, from our different experimental attempts, the ionization rate seems to be insensitive to the polarization of the $355\,$nm laser or its frequency shift on the order of tens of MHz using an AOM, but can be affected by the long-term frequency drift on a day-to-day basis. Specifically, by holding the $F$-type qubit Rabi rate at $2\pi\times 8\,$kHz, we observe that the typical timescale of the second ionization events fluctuates from about $500\,$ms to about $30\,$s on different days. Given the broad bandwitdh of the pulsed $355\,$nm laser of tens of GHz, the detuning to the relevant transitions may also be on the same order or even larger. Therefore, in the future it may be possible to suppress the second ionization events without sacrificing the manipulation speed the qubits by keeping only the central part of the spectrum to bridge the Raman transitions, while filtering out its long tails to avoid other undesired transitions. Also it may be desirable to use a $355\,$nm laser with a tunable central wavelength to scan and optimize the performance.
	
	%\bibliography{references}
	%merlin.mbs apsrev4-1.bst 2010-07-25 4.21a (PWD, AO, DPC) hacked
	%Control: key (0)
	%Control: author (0) dotless jnrlst
	%Control: editor formatted (1) identically to author
	%Control: production of article title (0) allowed
	%Control: page (1) range
	%Control: year (0) verbatim
	%Control: production of eprint (0) enabled
	%
	
\end{document}